\documentclass[prd,twocolumn,floatfix,amsmath,nofootinbib,amssymb,floatfix]{revtex4}
\usepackage{graphicx,color,dcolumn,booktabs,bm,multirow}
\usepackage{longtable,lscape}
\usepackage{txfonts}
\usepackage{overpic}
\usepackage{amssymb}
\usepackage{array}
\usepackage{indentfirst}
\usepackage{feynmf}   
\usepackage{slashed}  
\usepackage{cases}
\usepackage{color}
\usepackage{multirow}
\usepackage{epstopdf}
\usepackage{tabularx}
\usepackage{graphicx,color,dcolumn,booktabs,bm}
\usepackage[compat=1.1.0]{tikz-feynman}
\usepackage[colorlinks,
citecolor=blue,
anchorcolor=red,
menucolor=red,
linkcolor=red,
filecolor=red,
runcolor=red,
urlcolor=blue,
frenchlinks=red]{hyperref}

\begin{document}
	
 \title{$\psi(3S)$ and $\eta_c(3S)$ production in the $B^+$ decay processes}
\author{Zhuo Yu$^{1}$}
\author{Qi Wu$^{2}$}
\author{Man-Yu Duan$^{1,3}$}
\author{Zhou Rui$^4$\footnote{Corresponding author}}\email{jindui1127@126.com}
\author{Dian-Yong Chen$^{1,5}$\footnote{Corresponding author}}\email{chendy@seu.edu.cn}
\affiliation{$^1$ School of Physics, Southeast University, Nanjing 210094, China}
\affiliation{$^2$Institute of Particle and Nuclear Physics, Henan Normal University, Xinxiang 453007, China}
\affiliation{$^3$ School of Information Network Security, People's Public Security University of China, Beijing 100038, China }
\affiliation{$^4$ Department of Physics, Yantai University, Yantai 264005, China}
\affiliation{$^5$ Lanzhou Center for Theoretical Physics, Lanzhou University, Lanzhou 730000, China}
\date{\today}

\begin{abstract}
In the present work, we investigate the production of $\psi(4040)$ in the $B^+\to K^+ \psi(4040)$ process. Our estimations indicate that the unexpectedly large branching fraction could be well reproduced by taking both the short-range perturbative and long-range non-perturbative QCD contributions into account, where the long-range non-perturbative part plays an important role. Considering the structural similarity of $\eta_c(3S)$ and $\psi(3S)$, we apply the same model parameters to investigate the production of $\eta_c(3S)$ in the $B^+$ decay process, and the branching fraction is predicted  to be $(0.81_{-0.29}^{+0.31})\times 10^{-3}$. Based on the estimations in the present work and the relevant experimental measurements, we suggest that the experimentalists search for $\eta_c(3S)$ in the $J/\psi \omega$ invariant mass distributions of $B^+ \to J/\psi \omega K^+$ process at Belle II and LHCb Collaborations.                     
\end{abstract}
\maketitle

\section{introduction}
\label{sec:introduction}
Since the discovery of the $J/\psi$ meson~\cite{E598:1974sol,SLAC-SP-017:1974ind}, the charmonium spectroscopy has served as an ideal  laboratory for testing both perturbative and non-perturbative behaviors of Quantum Chromodynamics (QCD). The lower-lying charmonium states below the open-charm threshold can be well-described by conventional non-relativistic potential models, while the system becomes complicated as the mass moves above the threshold of a pair of charmed mesons. The properties of these excited states are influenced by the coupled-channel effects and non-perturbative long-distance dynamics, posing a long-standing challenge to both theoretical frameworks and experimental interpretation. A comprehensive review can be found in a recent review~\cite{Bai:2026atm}.

Among the production processes of charmonia, the $e^+e^-$ annihilation and the $B$ meson decay processes are particularly important. In the former process, the vector charmonia states can be produced directly, where the production of $\psi(nS)$ is characterized by the wave function at the origin. The higher radially excited states with more radial nodes exhibit a systematically suppressed $|\psi(0)|^2$, since their wave functions spread over a larger spatial region and reduce the probability amplitude at $r=0$, which results in smaller dilepton widths. For example, the dielectron widths of $J/\psi$, $\psi(2S)$, and $\psi(3S)$ are measured to be $(5.53\pm 0.10)$, $(2.33\pm 0.04)$, and $(0.86\pm 0.07)$ keV~\cite{ParticleDataGroup:2026mpi}, respectively, where $\psi(4040)$ is considered as the $\psi(3S)$ dominant charmonium state~\cite{Godfrey:1985xj, Barnes:2005pb, Eichten:2004uh, Li:2009zu, Bokade:2024tge, Wang:2022jxj, Man:2025vmm}, and we refer to $\psi(4040)$ as $\psi(3S)$ in the following. 

In the $B$ meson decay process, charmonia states with rich $J^{PC}$ quantum numbers could be produced, and abundant bottom mesons data samples have been collected by the LHCb and Belle/Belle II Collaborations, which makes $B$ meson decay one of the most important charmonium and charmonium-like state production processes, and an increasing number of higher excited charmonia have been observed in the $B$ meson decays (more details can be found in recent reviews~\cite{Hosaka:2016pey,Chen:2016qju,Esposito:2016noz,Chen:2016spr,Lebed:2016hpi,Guo:2017jvc,Olsen:2017bmm,Ali:2017jda,Liu:2019zoy,Brambilla:2019esw,Chen:2022asf,Meng:2022ozq,Wang:2025dur}). In addition, the charmonium production from $B$ meson decay serves as an ideal playground to disentangle short-distance perturbative QCD (pQCD) and long-distance non-perturbative QCD effects. So far the processes $B^+ \to K^+ J/\psi, K^+ \psi(2S)$, and $K^+ \psi(4040)$ have been experimentally observed and the observed branching fractions are collected in Table \ref{tab:br_BpsiK}. The investigation of $S$-wave vector charmonia production can provide an important test of the perturbative and non-perturbative effects in $B$ meson decays. 

\begin{table*}[htbp]
	\centering
	\caption{The branching fractions of $B^+\to \psi K^+$ decays.}
	\label{tab:br_BpsiK}
	\renewcommand\arraystretch{1.5}
	\setlength{\tabcolsep}{20pt}
	\begin{tabular}{cccc}
		\toprule[0.8pt] \toprule[0.8pt]
		& $B^+\to J/\psi K^+$ 
		& $B^+\to \psi(2S)K^+$ 
		& $B^+\to \psi(4040)K^+$ \\
		\midrule[0.8pt]
		Exp. (PDG)
		& $(1.020\pm0.019)\times10^{-3}$
		& $(6.24\pm0.20)\times10^{-4}$
		& $(1.6\pm0.5)\times10^{-3}$ \\
		Ref.~\cite{Chen:2005ht}
		& $(9.20^{+6.03}_{-7.99})\times 10^{-4}$ 
		& -
		& -
		\\
		Ref.~\cite{Zhang:2017fzl}
		& -
		& $(5.37^{+1.85}_{-2.22})\times 10^{-4}$
		& -
		\\
	\bottomrule[0.8pt]\bottomrule[0.8pt]
	\end{tabular}
\end{table*}

Generally, the charmonium production from $B^+$ meson decays is dominated by the short-distance contributions, where the $\bar{b}$ quark in the $B$ meson transits into a $\bar{c}$ quark by emitting a $W^+$ boson, and consequently, the $W^+$ boson decays into a $c$ quark and a $\bar{s}$ quark. The $c$ and $\bar{c}$ quarks hadronize as a charmonium, while the $\bar{s}$ and the $u$ quark in the initial $B^+$ meson compose a strange meson. This kind of process can be evaluated by pQCD approach, for example, in Ref.~\cite{Chen:2005ht}, the authors estimated the non-factorization corrections to naive factorization assumption in the pQCD approach, and the branching fraction of $B^+ \to J/\psi K^+$ was estimated to be  $(9.20^{+6.03}_{-7.99})\times 10^{-4}$, which is consistent with the experimental measurements, $(1.020 \pm 0.019)\times 10^{-3}$~\cite{ParticleDataGroup:2026mpi}. In addition, when including the next-to-next-to-leading order effects, the branching fraction of $B^+ \to \psi(2S) K^+ $ was estimated to be  $(5.37^{+1.85}_{-2.22})\times 10^{-4}$ in Ref.~\cite{Zhang:2017fzl}, which is also consistent with the experimental data, $(6.24\pm0.20)\times10^{-4}$~\cite{ParticleDataGroup:2026mpi}. In the pQCD approach, the wave function of the relevant particles will be involved in the convolution. Considering the nodes of the charmonia wave functions, the branching fractions are expected to be smaller for the higher excited charmonia. For $B^+ \to K^+ J/\psi$ and $B^+ \to K^+ \psi(2S)$, the experimental measurements and the pQCD estimations are consistent with such expectation. However, if we consider $\psi(4040)$ as $\psi(3S)$ dominant state, the measured branching fraction of $B^+ \to K^+ \psi(3S)$ is $(1.6\pm 0.5) \times 10^{-3}$, which is unexpectedly larger than those of $B^+ \to K^+ J/\psi$ and $B^+ \to \psi(2S) K^+ $. Whether the short-distance pQCD could explain such an anomalous phenomenon remains to be investigated.

In addition, the dominant $B^+$ meson decays occur in the following way: the $\bar{b}$ quark transits into a $\bar{c}$ by emitting a $W^+$ boson, and the $W^+$ boson decays into $c\bar{s}$, which further hadronizes as a charmed strange meson, while the  $\bar{c}$ and $u$ quarks hadronize as a charmed meson. This process is color enhanced compared with the direct charmonia production described above. Unlike $J/\psi$ and $\psi(2S)$, $\psi(4040)$ is above the $D^{(\ast)}\bar{D}^{(\ast)}$ thresholds, indicating that $\psi(4040)$ dominantly decays into a pair of charmed mesons. Thus, one can bridge the initial $B^+$ meson with the final charmonium and kaon by the charmed-(strange) meson loops, which could be considered as the phenomenological description of the long-distance non-perturbative QCD effects in the charmonium production. The meson loop contributions have been comprehensively discussed in charmonium production from bottom meson decay processes in Refs.~\cite{Cheng:2005wf, Cheng:2005bg, Cheng:2004ru, Pham:2003jp, Colangelo:2003sa}. As for $B^+ \to K^+ \psi(3S) $ process, these non-perturbative effects should be further clarified. Thus, in the present work, we investigate the process $B^+ \to K^+ \psi(3S)$ by considering both the short-range and long-range contributions, and check which part is dominant. In addition, the pseudoscalar partner of $\psi(3S)$, $\eta_c(3S)$, has not been observed yet. This state cannot be directly produced from $e^+ e^-$ annihilation process, and the $B$ meson decay becomes the most promising production process. Thus, in the present work, we can also predict the production of $\eta_c(3S)$ in bottom meson decays with the same mechanism. 

The rest of this work is organized as follows. After the introduction, we present the theoretical framework used in the estimation. The numerical results and relevant discussions are given in Section~\ref{Sec:Num}, and Section~\ref{Sec:Sum} is devoted to a brief summary.

\section{Theoretical framework}
\label{Sec:Method}

\begin{figure}[t]
	\begin{tabular}{cccc}
		\centering
\includegraphics[width=4.5cm]{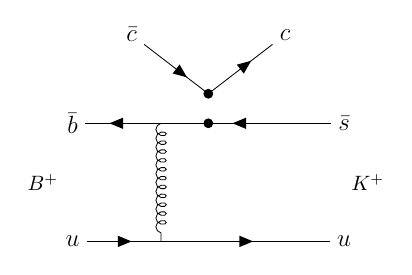}&
\includegraphics[width=4.5cm]{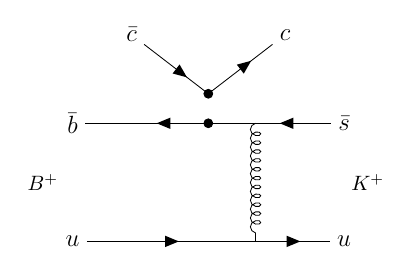}\\
		$(a)$ & $(b)$ \\ \\
	\includegraphics[width=4.5cm]{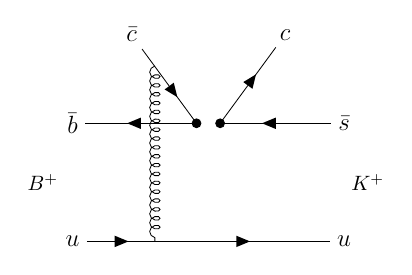}&
		\includegraphics[width=4.5cm]{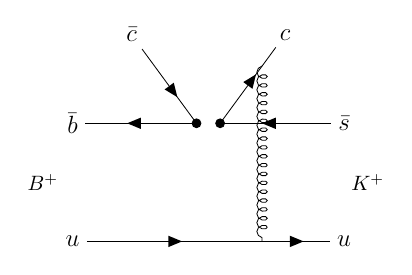}\\
		$(c)$ & $(d)$ \\ 
	\end{tabular}
	\caption{Typical short distance diagrams contributing to $B^+ \to K^+ \psi(3S)/\eta_c(3S)$ at the quark level.}\label{fig:feyndia}
\end{figure}

\subsection{Short-distance perturbative QCD estimations}
pQCD is one of the most popular factorization approaches for calculating the short-distance contributions to the bottom meson decays based on the \(k_T\) factorization theorem with \(k_T\) being the transverse momentum of partons~\cite{Li:1994cka}.
Once  \(k_T\) is included, the formalism is free from the endpoint singularities that appear in conventional collinear factorization. Sudakov resummation has also been introduced to effectively suppress long-distance contributions from regions of large transverse distance. The most important advantage of this method is that both factorizable and nonfactorizable contributions can be calculated self-consistently within the perturbative framework, without introducing additional non-perturbative parameters.


In our previous works~\cite{Rui:2015iia,Rui:2016opu,Rui:2017pre,Rui:2018fxz,Rui:2017fje,Rui:2023fiz}, the pQCD approach has been successfully applied to describe the decays of \(B\) and \(B_c\) mesons and the \(\Lambda_b\) baryon into the charmonium state. As of now, no pQCD calculations have been performed for the branching fraction of the $B^+ \to \psi(3S) K^+$  and $B^+ \to \eta_c(3S) K^+$  processes. In the present work, we investigate these two modes with the pQCD framework for the first time. The amplitude of the considered processes is written as 
\begin{eqnarray}
\mathcal{A} (B \to XK) &=& \mathcal F [V_{cb}^* V_{cs} a_2 - V_{tb}^* V_{ts} (a_3 + a_5 + a_7 + a_9)] \nonumber\\
&+& \mathcal M [V_{cb}^* V_{cs} C_2 - V_{tb}^* V_{ts} (C_4 - C_6 - C_8 + C_{10})],\nonumber\\
\end{eqnarray}
where $V_{ij}$ is the CKM matrix element. Here we define the combinations of the Wilson coefficients:
\begin{align*}
a_2 &= C_1+\frac{C_2}{3} + \frac{\alpha_s C_F}{4\pi N_c} C_2 (-18 + 12 \ln \frac{m_b}{\mu} + f_I), \\
a_i &= C_i+\frac{C_{i+1}}{3} + \frac{\alpha_s C_F}{4\pi N_c} C_{i+1} (-18 + 12 \ln \frac{m_b}{\mu} + f_I), \quad (i = 3, 9), \\
a_i &= C_i+\frac{C_{i+1}}{3} + \frac{\alpha_s C_F}{4\pi N_c} C_{i+1} (6 - 12 \ln \frac{m_b}{\mu} - f_I), \quad (i = 5, 6),
\end{align*}
with the color factor \(C_F = 4/3\) and the number of colors \(N_c = 3\).
The function $f_I$ arises from the vertex corrections, which are given in Ref.~\cite{Cheng:2001ez}. 
$\mathcal {F}$ and $\mathcal {M}$ describe the contributions from the factorizable and non-factorizable diagrams, respectively, shown in Fig.~\ref{fig:feyndia}. Their explicit factorization formulas can be written as 
\begin{align}\label{eq:f1}
\mathcal F &= -8 i \pi M^4  C_f f_{\eta}\int_0^1  d x_1 d x_3 \int_0^\infty b_1 b_3 d b_1 d b_3\phi_B(x_1, b_1) \notag\\ 
&\quad \Bigl\{\Bigl[ \left(r^2-1\right) \phi_a \left(\left(r^2-1\right)x_3-1\right)
-r_K \phi_p \notag\\
&\quad \left(r^2 \left(-2 x_3-1\right)+2 x_3-1\right)
+\left(r^2-1\right)\left(2x_3-1\right) r_K \phi_t \Bigr] \notag\\
&\quad \alpha_s(t_a) S_{ab}(t_a) h(\alpha_e, \beta_a, b_1, b_3) S_t(x_3)
\notag\\ 
&\quad + \Bigl[-\left(r^2-1\right) r^2 x_1 \phi _a-2 r_K \phi _p
   \left(r^2 \left(x_1+1\right)-1\right) \Bigr]\notag\\
&\quad \alpha_s(t_b) S_{ab}(t_b) h(\alpha_e, \beta_b, b_1, b_3) S_t(x_1)\Bigr\},
\end{align}
\begin{align}\label{eq:m1}
\mathcal M &= \frac{16 i \sqrt{2} \pi  M^4 C_f }{\sqrt{N_c}} \int_0^1  d x_1 d x_2 d x_3 \int_0^\infty b_1 b_2 d b_1 d b_2 \notag\\ 
&\quad \phi_B(x_1, b_1)\Bigl\{ \Bigl[ \Bigl(r r_c
   \psi _s \left(\left(r^2-1\right) \phi _a+4 r_K
   \phi _p\right)  \notag\\ 
&\quad+\psi _v \Bigl(\left(r^4-1\right)
   \left(x_1+x_2-1\right) \phi _a \notag\\
&\quad +r^2 r_K
   \left(\left(x_1+2 x_2+x_3-2\right) \phi
   _p+\left(x_1-x_3\right) \phi _t\right)\notag\\
&\quad+x_3 r_K
   \left(\phi _t-\phi
   _p\right)\Bigr)\Bigr) \Bigr]  \alpha_s(t_c) S_{cd}(t_c) h(\alpha_e, \beta_c, b_1, b_2) 
\notag\\ 
&\quad + \Bigl[\Bigl(r r_c
   \psi _s \left(r^2 \left(-\phi _a\right)+\phi _a-4
   r_K \phi _p\right) \notag\\
&\quad-\psi _v \Bigl(x_1
   \left(\left(r^2-1\right) \phi _a+r^2 r_K
   \left(\phi _p-\phi
   _t\right)\right)+\left(r^2-1\right) \notag\\
&\quad x_3
   \left(\left(r^2-1\right) \phi _a+r_K \left(\phi
   _p+\phi _t\right)\right)\notag\\
&\quad +x_2 \left(r^4 \left(-\phi
   _a\right)+\phi _a-2 r^2 r_K \phi
   _p\right)\Bigr)\Bigr) \Bigr]\notag\\
&\quad \alpha_s(t_d) S_{cd}(t_d) h(\alpha_e, \beta_d, b_1, b_2) \Bigr\},
\end{align}
for the $\eta_c(3S)$ mode and 
\begin{align}\label{eq:f2}
\mathcal F &= -8 \pi M^4  C_f f_{\psi}\int_0^1  d x_1 d x_3 \int_0^\infty b_1 b_3 d b_1 d b_3\phi_B(x_1, b_1) \notag\\ 
&\quad \Bigl\{\Bigl[   \left(r^2-1\right) \phi _a \left(\left(r^2-1\right)
   x_3-1\right)+\left(r^2-1\right) \left(2 x_3-1\right) r_K \phi _p\notag\\
&\quad-r_K \left(r^2 \left(-2
   x_3-1\right)+2 x_3-1\right) \phi _t  \Bigr] \notag\\
&\quad \alpha_s(t_a) S_{ab}(t_a) h(\alpha_e, \beta_a, b_1, b_3) S_t(x_3)
\notag\\ 
&\quad + \Bigl[\left(\left(r^2-1\right) r^2 x_1 \phi _a+2 r_K \phi _p \left(r^2
   \left(x_1-1\right)+1\right)\right) \Bigr]\notag\\
&\quad \alpha_s(t_b) S_{ab}(t_b) h(\alpha_e, \beta_b, b_1, b_3) S_t(x_1)\Bigr\},
\end{align}
\begin{align}\label{eq:m2}
\mathcal M &= \frac{16 \sqrt{2} \pi  M^4 C_f }{\sqrt{N_c}} \int_0^1  d x_1 d x_2 d x_3 \int_0^\infty b_1 b_2 d b_1 d b_2 \notag\\ 
&\quad \phi_B(x_1, b_1)\Bigl\{ \Bigl[   r \left(r^2-1\right) \phi _a r_c \psi _t \notag\\
&\quad+\psi _L \Bigl(\left(r^2-1\right)^2 \left(x_1+x_2-1\right) \phi _a \notag\\
&\quad+r^2 r_K \left(\left(x_1-x_3\right) \phi
   _p+\left(x_1+2 x_2+x_3-2\right) \phi _t\right) \notag\\
&\quad+x_3 r_K \left(\phi _p-\phi
   _t\right)\Bigr)  \Bigr]  \alpha_s(t_c) S_{cd}(t_c) h(\alpha_e, \beta_c, b_1, b_2) 
\notag\\ 
&\quad + \Bigl[ -\left(r^2-1\right) \Bigl(-r \phi _a r_c \psi _t\notag\\
&\quad +\phi _a
   \psi _L \left(r^2 x_2-r^2 x_3+x_2+x_3\right)-x_3 r_K \psi _L \phi _p\Bigr)\notag\\
&\quad+x_1 \psi _L
   \left(\left(r^2-1\right) \phi _a+r^2 r_K \left(\phi _t-\phi _p\right)\right)\notag\\
&\quad+r_K \phi _t \left(4 r
   r_c \psi _t+r^2 \left(x_3-2 x_2\right) \psi _L-x_3 \psi _L\right)  \Bigr]\notag\\
&\quad \alpha_s(t_d) S_{cd}(t_d) h(\alpha_e, \beta_d, b_1, b_2) \Bigr\},
\end{align}
for the $\psi(3S)$ mode. We use $x_i$ to denote the momentum
fraction of the quark in each meson, and $b_i$ to denote
the conjugate space coordinate of the corresponding transverse
momentum. $r_{(c)}$ is the mass ratio between the charmonium state (charm quark) and the bottom meson.  The mass ratio \(r_K = m_0^K/m_B\) with \(m_0^K\) being the chiral scale associated with the kaon.
Here, $\psi_{v,s,L,t}$ and $\phi_{a,p,t}$ represent the light-cone distribution amplitudes (LCDAs) for the charmonia and $K$ meson, respectively. 
$\phi_B(x_1, b_1)$ is the model of $B$ meson wave function in the impact parameter $b$ space.
The explicit expressions for the Sudakov factor $S(t)$, the jet function $S_t(x)$, the hard function $h$, and the scales $t_{a,b,c,d}$ are collected in Appendix~\ref{Sec:AppendixC}.


With the above formulas and inputs, we derive the branching fractions
\begin{eqnarray}
	\mathcal{B}[B^+ \rightarrow \psi(3S) K^+]& =& 2.4^{+0.7}_{-0.5}\times 10^{-4}, \nonumber\\
  \mathcal{B}[B^+ \rightarrow \eta_c(3S) K^+] &=& 3.1^{+1.1}_{-0.7}\times 10^{-4}, 
 \end{eqnarray}
where the error comes from the non-perturbative parameter in the $B$-meson wave function. The short distance pQCD estimation for $B^+ \to K^+ \psi(3S)$ is much smaller than the measured one as shown in Table~\ref{tab:br_BpsiK}. This large discrepancy between short range perturbative QCD estimations and the experimental data indicates the long-distance non-perturbative QCD effects should play an important role in understanding the anomalous large branching fraction of $B^+ \to K^+ \psi(3S)$, which will be discussed in the following subsection.


\begin{figure}[t]
	\begin{tabular}{cccc}
		\centering
\includegraphics[width=4.2cm]{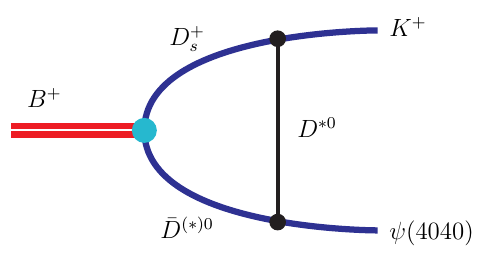}&
		\includegraphics[width=4.2cm]{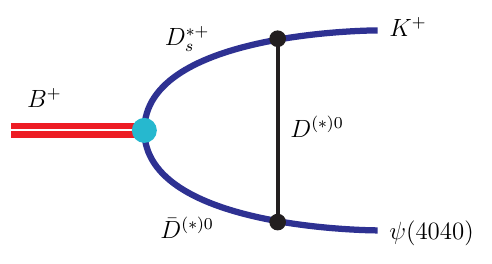}\\
		$(a)$ & $(b)$ \\ \\
		\includegraphics[width=4.2cm]{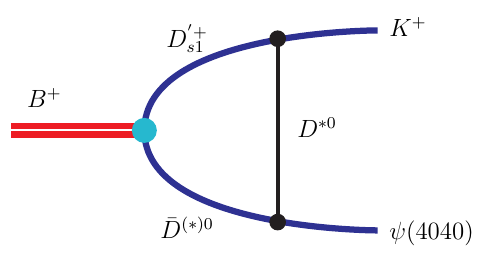}&
		\includegraphics[width=4.2cm]{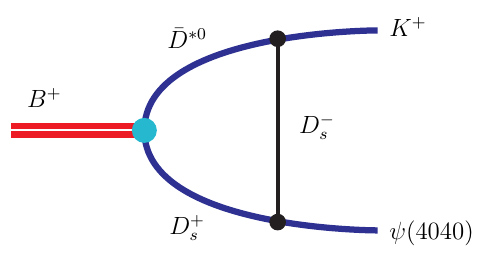}\\
		$(c)$ & $(d)$ \\ 
	\end{tabular}
	\caption{Typical diagrams contributing to $B^+ \to K^+ \psi(4040)$ at the hadron level.}\label{fig:FeynDiagram1}
\end{figure}

\begin{table}[tb]
	\caption{Branching fractions of the primary $B^+$ weak decays utilized in the present work. Experimental values are taken from the PDG, alongside our theoretical estimates derived from the weak decay framework. \label{Tab:BtoDD}}
	\renewcommand\arraystretch{1.5}
	\setlength{\tabcolsep}{8pt}
	\begin{tabular}{ccc}
		\toprule[1.2pt]
		Process & PDG Value~\cite{ParticleDataGroup:2026mpi} & Theoretical estimations \\
		\midrule[1pt]
		$B^+\to D_{s}^+\bar{D}^0$         & $(9.0\pm0.9)\times10^{-3}$  & $14.75\times10^{-3}$ \\ 
		$B^+\to D_{s}^{*+}\bar{D}^{0}$     & $(7.6\pm1.6)\times10^{-3}$  & $11.75\times10^{-3}$ \\
		$B^+\to D_{s}^+\bar{D}^{*0}$        & $(8.2\pm1.7)\times10^{-3}$  & $7.79\times10^{-3}$  \\ 
		$B^+\to D_{s}^{*+}\bar{D}^{*0}$    & $(1.71\pm0.24)\%$           & $23.9\times10^{-3}$  \\
		$B^+\to D_{s1}^{\prime+}\bar{D}^0$   & $(3.1_{-0.9}^{+1.0})\times10^{-3}$ & $1.91\times10^{-3}$ \\ 
		$B^+\to D_{s1}^{\prime +}\bar{D}^{*0}$ & $(1.20\pm0.30)\%$           & $9.69\times10^{-3}$  \\
		\bottomrule[1.2pt]
	\end{tabular}
\end{table}

\subsection{Long-distance non-perturbative QCD effects}
As we mentioned in the introduction, the bottom meson dominantly decays into a charmed and a charmed-strange meson by weak interaction, and this anti-charmed and charmed-strange meson pair can transit into $\psi(3S)$  and a kaon via exchanging a proper charmed(-strange) meson. Here, we use $\mathcal{H}_W$ and $\mathcal{H}_T$ to refer to the initial weak decay and the subsequent hadronic transition interactions, respectively, then, the total decay matrix element can be factorized as
\begin{eqnarray}
	&& \langle K^+ \psi(4040) \left|\mathcal{H}_T \mathcal{H}_W \right|B^+\rangle \nonumber\\ 
	&&\qquad = \sum_{M_1 M_2} \langle K^+ \psi(4040) \left|\mathcal{H}_T \right| M_1 M_2 \rangle \langle M_1 M_2 \left| \mathcal{H}_W \right|B^+\rangle, \label{eq:amplitude_factorization}\nonumber \\
\end{eqnarray}
where $M_1 M_2$ denotes the intermediate charmed and charmed-strange meson pairs contributing to the transition. In principle, all kinematically accessible intermediate anti-charmed and charmed-strange mesons that can connect the initial $B^+$ and final $K^+ \psi(4040)$ states should be included. In the present estimations, only the dominant channels are taken into account, such as $D_s^{(*)+} \bar{D}^{(*)0}$ and $D_{s1}^{\prime}(2460)^+ \bar{D}^{(*)0}$, since the branching fractions of corresponding $B$ decays are relatively large. In Table \ref{Tab:BtoDD}, we present the measured branching fractions of the $B^+$ decay process involved in the present estimation~\cite{ParticleDataGroup:2026mpi}. Considering that $\psi(3S)$ dominantly decays into $D^{(\ast)}\bar{D}^{(\ast)}$ and $D_s^+ D_s^-$,  we can construct the meson loops contributing to $B^+ \to K^+ \psi(3S)$, and the typical diagrams are presented in Fig.~\ref{fig:FeynDiagram1}. The explicit meson loop diagrams involved in the present estimations are collected in Table~\ref{Tab:all loop}. With the same mechanism, we present the typical diagram and explicit meson loop contributing to $B^+ \to K^+ \eta_c(3S)$ are presented in Fig.~\ref{fig:FeynDiagram2} and Table~\ref{Tab:all loop}.

\begin{table}[tb]
	\caption{Explicit meson loops contributing to $B^+\rightarrow K^+\psi(3S)$ and $B^+\rightarrow K^+\eta_c(3S)$ in the present work.  Here, the mesons enclosed within the parentheses represent the intermediate states produced in the primary $B$ decay, while the meson outside denotes the exchanged degree of freedom.}
	\label{Tab:all loop}
	\renewcommand\arraystretch{1.5}
	\setlength{\tabcolsep}{10pt}
	\begin{tabular}{cc}
		\toprule[1pt]
		Process  & Loops  \\
		\midrule[1pt]
		$B^+\rightarrow K^+\psi(3S)$  & $[D_s^{+} \bar{D}^{0}] D^{*0}$,~ $[D_s^{*+} \bar{D}^{0}] D^{0}$,~ $[D_s^{*+} \bar{D}^{0}] D^{*0}$ \\
	  & $[D_s^{+} \bar{D}^{*0}] D^{*0}$,~ $[D_s^{*+} \bar{D}^{*0}] D^{0}$,~ $[D_s^{*+} \bar{D}^{*0}] D^{*0}$  \\
	  & $[D_{s1}^{+} \bar{D}^{*0}] D^{*0}$,~ $[D_{s1}^{+} \bar{D}^{0}] D^{*0}$,~ $[\bar{D}^{*0} D_{s}^{+}] D_{s}^{-}$  \\
        \midrule[1pt]
		$B^+\rightarrow K^+\eta_c(3S)$ & $[D_s^{+} \bar{D}^{0}] D^{*0}$,~  $[D_s^{*+} \bar{D}^{0}] D^{*0}$ \\
	  & $[D_s^{+} \bar{D}^{*0}] D^{*0}$,~ $[D_s^{*+} \bar{D}^{*0}] D^{0}$,~ $[D_s^{*+} \bar{D}^{*0}] D^{*0}$  \\
	  & $[D_{s1}^{+} \bar{D}^{*0}] D^{*0}$,~ $[D_{s1}^{+} \bar{D}^{0}] D^{*0}$,  \\
		\bottomrule[1pt]
	\end{tabular}
\end{table}


In the present work, all the processes are evaluated at the hadronic level and the effective Lagrangians are employed to describe the interactions between the involved hadrons. As for the $B^+$ weak decay vertex,  we adopt the parametrized hadronic matrix elements obtained by the effective Hamiltonian at the quark level, which are~\cite{Cheng:2003sm,Soni:2021fky}, 
\begin{eqnarray}
\label{eq:weak transition}
&&\left\langle0|J_\mu|P(p_1)\right\rangle = -if_p p_{1\mu}\;,\nonumber\\
&&\left\langle0|J_\mu|V(p_1,\epsilon)\right\rangle =f_V \epsilon_\mu m_V \;,\nonumber\\
&&\left\langle P(p_2)|J_\mu|B(p_0)\right\rangle \nonumber\\&&\quad=\left[P_\mu-\frac{m^2_{B}-m^2_P}{q^2}q_\mu\right]F_1\left(q^2\right) 
+\frac{m^2_{B}-m^2_P}{q^2}q_\mu F_0\left(q^2\right) \;,\nonumber\\
&&\left\langle V(p_2,\epsilon)|J_\mu|B(p_0)\right\rangle \nonumber\\
&&\quad=\frac{i\epsilon^\nu}{m_{B}+m_V}\Bigg\{i\varepsilon_{\mu\nu\alpha\beta}P^\alpha q^\beta A_V\left(q^2\right)\nonumber\\
&&\quad -(m_{B}+m_V)^2 g_{\mu\nu}A_1\left(q^2\right) +P_\mu P_\nu A_2\left(q^2\right)\nonumber\\
&&\quad+2m_V(m_{B}+m_V)\frac{P_\nu q_\mu}{q^2}\left[A_3\left(q^2\right)-A_0\left(q^2\right)\right] \Bigg\}\;,
\end{eqnarray}
with $J_{\mu}=\bar{q}_1 \gamma_{\mu}(1-\gamma_5)q_2$, $P_{\mu}=p_{0\mu}+p_{2\mu}$ and $q_{\mu}=p_{0\mu}-p_{2\mu}$.  $A_{0,1,2,V,+,-}\left(q^2\right)$ and $F_{0,1,+,-}\left(q^2\right)$ are the weak transition form factors, while $A_3\left(q^2\right)$ is the linear combination of form factors $A_1\left(q^2\right)$ and $A_2\left(q^2\right)$, which is~\cite{Cheng:2003sm},
\begin{eqnarray}
A_3\left(q^2\right)=\frac{m_B+m_{V}}{2m_{V}}A_1\left(q^2\right)-\frac{m_B-m_{V}}{2m_{V}}A_2\left(q^2\right). 
\end{eqnarray}

The amplitudes of the $B^+$ weak transition could be constructed by the products of two hadronic matrix elements. Here, we take the vertex of $B^+ \rightarrow D_{s1}^{\prime +}\bar{D}^*$ as an example. The amplitude reads,
\begin{eqnarray}
&&\mathcal{A}(B^+ \to D_{s1}^{\prime+} \bar{D}^*) =	\mathcal{A}_{\mu\nu }^{B^+ \to D_{s1}^{\prime+} \bar{D}^*} \epsilon_{D_{s1}^{\prime}}^{\mu} \epsilon_{\bar{D}^\ast}^\nu \nonumber\\
&&\qquad = \frac{G_F}{\sqrt{2}}V_{cb}^*V_{cs}a_1	\left \langle D_{s1}^{\prime+} \left | J_\mu^\dagger \right |0  \right \rangle 	\left \langle \bar{D}^* \left | J_\mu \right |B^+  \right \rangle ,
\end{eqnarray}
where $G_F$ is the Fermi constant, $V_{cb}^*$ and $V_{cs}$ are the CKM matrix elements. $a_1=c_1^{eff}+c_2^{eff}/N_c$ with $c_{1,2}^{eff}$ being effective Wilson coefficients obtained by the factorization approach~\cite{Bauer:1986bm}. In addition, we adopt $G_F=1.166 \times 10^{-5} {\rm GeV}^{-2}$, $V_{cb}=0.041$, $V_{cs}=0.987$ and $a_1=1.05$ as those in Refs.~\cite{ParticleDataGroup:2020ssz,Ali:1998eb,Ivanov:2006ni} in the following estimations.

Besides the weak transition, the interactions of charmonia and heavy-light mesons could also be described by the effective Lagrangians constructed in the heavy quark limit, and the relevant effective Lagrangians read,
\begin{eqnarray}
\label{eq:charmonia DD}
&&\mathcal{L}_{\psi \mathcal{D}^{(*)} \mathcal{D}^{(*)}} = -i g_{\psi \mathcal{D} \mathcal{D}} \psi_{\mu} \mathcal{D}^{\dagger} \overleftrightarrow{\partial^{\mu}} \mathcal{D} \nonumber\\
&&\qquad+g_{\psi \mathcal{D}^{*} \mathcal{D}} \epsilon^{\mu \nu \alpha \beta} \partial_{\mu} \psi_{v}\left(\mathcal{D}_{\alpha}^{*} \overleftrightarrow{\partial_{\beta}} \mathcal{D}^{\dagger}-\mathcal{D} \overleftrightarrow{\partial_{\beta}} \mathcal{D}_{\alpha}^{* \dagger}\right) \nonumber\\
&&\qquad+i g_{\psi \mathcal{D}^{*} \mathcal{D}^{*}} \psi^{\mu}\left(\mathcal{D}_{v}^{*} \overleftrightarrow{\partial^{*}} \mathcal{D}_{\mu}^{* \dagger}+\mathcal{D}_{\mu}^{*} \overleftrightarrow{\partial} \mathcal{D}_{v}^{* \dagger}\right. \left.-\mathcal{D}_{v}^{*} \overleftrightarrow{\partial_{\mu}} \mathcal{D}^{* \nu \dagger}\right)\nonumber\\
&&\mathcal{L}_{\eta_c \mathcal{D}^{*} \mathcal{D}^{(*)}} = -i g_{\eta_c \mathcal{D}^* \mathcal{D}}\eta_c \left( \mathcal{D}\overleftrightarrow{\partial_{\mu}}\mathcal{D}^{*\dagger\mu}+\mathcal{D}^{*\mu}\overleftrightarrow{\partial_{\mu}}\mathcal{D}^\dagger \right) \nonumber\\
&&\qquad-g_{\eta_c \mathcal{D}^* \mathcal{D}^*}\epsilon^{\mu \nu \alpha \beta}\partial_{\mu}\eta_c \mathcal{D}_{\nu}^*\overleftrightarrow{\partial_{\alpha}}\mathcal{D}_{\beta}^{*\dagger} ,
\end{eqnarray}
where  $\mathcal{D}^{(*)}=(D^{0(*)}, D^{+(*)}, D_s^{+(*)})$  and $A\overleftrightarrow{\partial_{\mu}}B = A \partial^\mu B - B \partial^\mu A$.

Similar to the coupling between charmonia and heavy-light mesons, the coupling between light pseudoscalar states and the charmed(-strange) meson pair can be constructed by considering the heavy quark limit and chiral symmetry~\cite{Casalbuoni:1996pg,Casalbuoni:1992gi,Casalbuoni:1992dx}, and the relevant effective Lagrangians are,
\begin{eqnarray}
\label{eq:D(1)DP}
&&\mathcal{L}_{\mathcal{D}^*\mathcal{D}^*\mathcal{P}} = g_{\mathcal{D}^*\mathcal{D}^*\mathcal{P}}\varepsilon_{\mu\nu\alpha\beta}\left(\mathcal{D}_b^{*\mu}\overleftrightarrow{\partial^{\alpha}}\mathcal{D}_a^{*\nu\dagger}\right)\partial^{\beta}\mathcal{P}_{ba},\nonumber\\
&&\mathcal{L}_{\mathcal{D}_1^{\prime}\mathcal{D}^{*}\mathcal{P}}=g_{\mathcal{D}_1^{\prime}\mathcal{D}^{*}\mathcal{P}}\Big[\mathcal{D}_{1b\mu}^\prime\stackrel{\leftrightarrow}{\partial^{\mu}}\mathcal{D}_{a\nu}^{*\dagger}\partial^\nu\mathcal{P}_{ba}\nonumber\\
&&\qquad -\mathcal{D}_{1b\mu}^\prime\stackrel{\leftrightarrow}{\partial^{\nu}}\mathcal{D}_{a}^{*\mu\dagger}\partial_\nu\mathcal{P}_{ba}+\mathcal{D}_{1b\mu}^\prime \stackrel{\leftrightarrow}{\partial^{\nu}} \mathcal{D}_{\nu}^{*\dagger} \mathcal{P}_{ba} \Big], \qquad
\end{eqnarray}
with $\mathcal{D}_1=(D_1^0, D_1^+, D_{s1}^+)$, and $\mathcal{P}$ is the matrix form of the light pseudoscalar mesons and its concrete form is,
\begin{eqnarray}\label{eq:pscalar matrix}
	\mathcal{P} &=&
	\left(\begin{array}{ccc}
		\frac{\pi^{0}}{\sqrt 2}+\alpha\eta+\beta\eta^\prime &\pi^{+} &K^{+}\\
		\pi^{-} &-\frac{\pi^{0}}{\sqrt2}+\alpha\eta+\beta\eta^\prime&K^{0}\\
		K^{-} &\bar K^{0} &\gamma\eta+\delta\eta^\prime
	\end{array}\right).
\end{eqnarray}


With the above effective Lagrangians, we can obtain the amplitudes corresponding to the meson loop contributions to the considered processes. Here, we take $B^+ \to (D_s^+ \bar{D}^0) D^{*0} \to K^+ \psi(3S)$ as an example. The corresponding amplitude reads,
\begin{eqnarray}
\label{Eq:Amp}
&&\mathcal{M}_{(D_s^+ \bar{D}^0) D^{*0}}= i^3 \int\frac{d^4q}{(2\pi)^4}\Big[\mathcal{A}^{B\rightarrow D_{s} \bar{D}^{0}}(p_1,p_2)\Big]\Big[-g_{\mathcal{D}^*\mathcal{D}\mathcal{P}}(i p_{3\beta})\Big] \nonumber\\
&&\qquad\times \Big[i g_{\psi \mathcal{D}^*\mathcal{D}}\varepsilon_{\mu_1\delta\phi\beta_1}p_4^{\mu_1}(p_2-q)^{\beta_1}\epsilon^\delta(p_4)\Big] \nonumber\\
&&\qquad\times\Big[\frac{1}{p_1^2-m_1^2}\Big]\Big[\frac{1}{p_2^2-m_2^2}\Big]\Big[\frac{-g^{\beta\phi}+q^{\beta} q^{\phi}/m_q^2}{q^2-m_q^2}\Big]\mathcal{F}^2(q^2,m_q^2). \nonumber\\
\end{eqnarray}
The rest of the amplitudes contributing to $B^+ \to K^+ \psi(3S)$ are collected in Appendix~\ref{Sec:AppendixA}, while the amplitudes relevant to $B^+ \to K^+ \eta_c(3S)$ are listed in Appendix~\ref{Sec:AppendixB}. In addition, the momentum of the involved heavy-light meson can be expressed as $p^\mu = m_Q v^\mu + k_\mu$ with $v$ being the heavy meson velocity and $k$ being the residual momentum. The mass of heavy-light meson can approximately be $m_Q+m_q$. Thus in the heavy quark limit, i.e., $m_Q \to \infty$, the numerator of the heavy-light meson propagator, $-g^{\mu\nu}+p^\mu p^\nu/m^2$, can be simplified to $-g^{\mu\nu}+v^\mu v^\nu$. 

It is worth noting that a form factor $\mathcal{F}(q^2,m^2)$ is introduced to compensate for the off-shell effects, internal structure of involved mesons, and regulate the ultraviolet divergence in the loop integrals as well. In the present estimations, we take a form factor in the monopole form, which is,
\begin{equation}
	\mathcal{F}\left(q^2,m^2\right) = \frac{m^2-\Lambda^2}{q^2-\Lambda^2},
\end{equation}
where $m$ and $q$ are the mass and the momentum of the exchanged meson, respectively. The cutoff parameter $\Lambda$ could be further reparameterized as ${ \Lambda} = m + \alpha {\rm \Lambda_{QCD}}$ with $\Lambda_{\rm QCD}$ = 220 MeV, where the parameter $\alpha$ is expected to be of order unity ~\cite{Cheng:2004ru}. 

With the amplitudes in Eq.~\eqref{Eq:Amp} and Appendix~\ref{Sec:AppendixA}, we can obtain the total amplitude for $B^+ \to K^+ \psi(3S)$, and then the partial width can be estimated by,
\begin{eqnarray}
	&&\Gamma_{B^+ \to K^+ \psi/\eta_c} = \frac{1}{8\pi} \frac{|\vec{p}\,|}{m_{B^+}^2}\overline{\left|\mathcal{M}_{B^+\to K^+\psi/\eta_c}\right|^2} ,
\end{eqnarray}
where $|\vec{p}|$ is the magnitude of the final-state momentum in the rest frame of the $B^+$, and the overline above the amplitude indicates the sum over the spins of final states. In the same way, we can estimate the partial width of $B^+ \to K^+ \eta_c(3S)$.


\section{Numerical Results and Discussions}
\label{Sec:Num}
\subsection{Coupling Constants}
Before presenting the numerical results, some relevant parameters should be clarified. The $f_P$ and $f_V$ in Eq.~\eqref{eq:weak transition} are the decay constants of the charmed strange mesons, and their values in the present work are taken as $f_{D_s}=250$ MeV, $f_{D_s^*}=272$ MeV and $f_{D_{s1}^\prime}=158$ MeV as in Refs.~\cite{FlavourLatticeAveragingGroup:2019iem,Hwang:2004kga}. In addition, the form factors in Eq.~\eqref{eq:weak transition} are usually estimated in the quark model and are known only in
the spacelike region. Some methods like analytic continuation could cover the timelike region where the physical decay processes occur. In Refs.~\cite{Cheng:2003sm,Soni:2021fky}, the form factors are parameterized in the form,  
\begin{equation}\label{eq:FQzeta}
	F\left(q^2\right) = \frac{F(0)}{1-a\zeta+b\zeta^2},
\end{equation}
with $\zeta = Q^2/m_{B}^2$. The relevant parameters $F(0)$, $a$, and $b$ for each form factor are collected in Table~\ref{Tab:F0ab}. In order to avoid ultraviolet divergences in the loop integrals and to evaluate them via Feynman parameterization, we further parameterize these form factors in the form,
\begin{equation}\label{eq:FQlambda}
	F\left(q^2\right) = F(0)\frac{\Lambda_1^2}{Q^2-\Lambda_1^2}\frac{\Lambda_2^2}{Q^2-\Lambda_2^2}.
\end{equation}
By fitting Eq.~\eqref{eq:FQlambda} with Eq.~\eqref{eq:FQzeta}, we can obtain the values of $\Lambda_1$ and $\Lambda_2$ for each form factor, which are listed in Table~\ref{Tab:L1L2}. To test the reliability of the above parameters, we estimate the branching fractions of the involved bottom meson decays, and the estimated results from Eq.~\eqref{eq:weak transition} are presented in Table~\ref{Tab:BtoDD}. From the table, one can find that the theoretical estimations are generally consistent with the experimental measurements within uncertainties.

\begin{table}
	\caption{\label{Tab:F0ab}The values of the parameters $F(0)$, $a$ and $b$ in the form factors~\cite{Cheng:2003sm}.}
	\renewcommand\arraystretch{1.5}
	\setlength{\tabcolsep}{20pt}
	\begin{tabular}{cccc}
		\toprule[1pt]
		Parameter &$F(0)$ &$a$ &$b$ \\
		\midrule[1pt]
		$F_0$ &0.67 &0.65 &0.00 \\
		$F_1$ &0.67 &1.25 &0.39 \\
		$A_0$ &0.64 &1.30 &0.31 \\
		$A_1$ &0.63 &0.65 &0.02 \\
		$A_2$ &0.61 &1.14 &0.52 \\
		$V$ &0.75 &1.29 &0.45 \\
		\bottomrule[1pt]	 
	\end{tabular}
\end{table}

\begin{table}
	\caption{\label{Tab:L1L2}Values of the parameters $\Lambda_1$ and $\Lambda_2$ obtained by fitting Eq.~\eqref{eq:FQlambda} with Eq.~\eqref{eq:FQzeta}.}
	\renewcommand\arraystretch{1.5}
	\setlength{\tabcolsep}{8pt}
	\begin{tabular}{ccccccc}
		\toprule[1pt]
		Parameter &$F_{0}$ &$F_{1}$ &$A_{0}$ &$A_{1}$ &$A_{2}$ &$V$ \\
		\midrule[0.8pt]
		$\Lambda_1$ &7.75 &6.53 &6.50 &8.95 &6.65 &6.30 \\
		$\Lambda_2$ &11.00 &6.84 &7.10 &8.75 &7.55 &7.10 \\
		\bottomrule[1pt]
	\end{tabular}
\end{table}

Considering the chiral symmetry and heavy quark limit, the coupling constants of charmed meson with pseudoscalar meson can be related to some gauge coupling by~\cite{Isola:2003fh,Falk:1992cx, Casalbuoni:1996pg},
\begin{eqnarray}
	&&g_{\mathcal{D}\mathcal{D}^*\mathcal{P}} =-\frac{2g}{f_{\pi}}\sqrt{m_{\mathcal{D}}m_{\mathcal{D}^*}}, \nonumber\\
	&&g_{\mathcal{D}^*\mathcal{D}^*\mathcal{P}} = \frac{2g}{f_{\pi}}\frac{1}{\sqrt{m_{\mathcal{D}^*}m_{\mathcal{D}^*}}}, \nonumber\\
	&&g_{\mathcal{D}_1^\prime \mathcal{D}^*\mathcal{P}} = \frac{2h}{f_{\pi}},
\end{eqnarray}
with $g=0.59$, $f_{\pi}=0.132~{\rm GeV}$~\cite{Chen:2019asm,Liu:2011xc,Isola:2003fh,Falk:1992cx} and $h=-0.56$ \cite{Casalbuoni:1996pg}.

The coupling constants between the $\psi(3S)/\eta_c(3S)$ and the charmed meson pairs are determined from the partial decay widths calculated within the $^3P_0$ model~\cite{Barnes:2005pb}, as listed in Table~\ref{Tab:Barnes}. Specifically, by utilizing the effective Lagrangians in Eq.~\eqref{eq:charmonia DD}, we extract the numerical values of the coupling constants as
\begin{eqnarray}
	&&g_{\psi(3S) D D} = 0.18,~ g_{\psi(3S) D^* D} = 0.64, \nonumber\\
	&&g_{\psi(3S) D^* D^*} = 7.98,~ g_{\psi(3S) D_s D_s} = 1.87, \nonumber\\
	&&g_{\eta_c(3S) D^* D} = 2.46,~ g_{\eta_c(3S) D^* D^*} = 1.92.
\end{eqnarray}

\begin{table}[tb]
	\caption{The open charm decay widths of $\psi(3S)$ and $\eta_c(3S)$ calculated with $^3P_0$ model~\cite{Barnes:2005pb}.}
	\label{Tab:Barnes}
	\renewcommand\arraystretch{1.5}
	\setlength{\tabcolsep}{15pt}
	\begin{tabular}{cccc}
		\toprule[1pt]
		Meson & State & Mode & $\Gamma_{thy}({\rm MeV)}$  \\
		\midrule[1pt]
		$\psi(4040)$ & $3^3S_1$ & $DD$ & 0.1 \\
	      &  & $DD^*$ & 33 \\
		  &  & $D^*D^*$ & 33 \\
            &  & $D_sD_s$ & 7.8 \\
          \midrule[1pt]
		$\eta_c(4043)$ & $3^3S_0$ & $DD^*$ & 47 \\
		 &  & $D^*D^*$ & 33 \\
		\bottomrule[1pt]
	\end{tabular}

\end{table}

\subsection{Branching Fractions}
With the above preparations, all the relevant parameters and coupling constants have been determined except for the model parameter $\alpha$ introduced by the form factor in Eq.~\eqref{eq:FQlambda}, which is typically expected to be of order unity as indicated in Ref.~\cite{Cheng:2004ru}. In the present work, we vary $\alpha$ from 1 to 4 to investigate the parameter dependence of the branching fractions for $B^+ \to K^+ \psi(3S)$. By matching the present estimations with the experimental measurements, we can determine the proper $\alpha$ range, and discuss whether the parameter is reasonable. In addition, this constrained parameter range can be further applied to predict the $B^+ \to K^+\eta_c(3S)$ process due to the structural similarity of $\eta_c(3S)$ and $\psi(3S)$.

\begin{figure}[t]
	\centering
 \includegraphics[width=8.5cm]{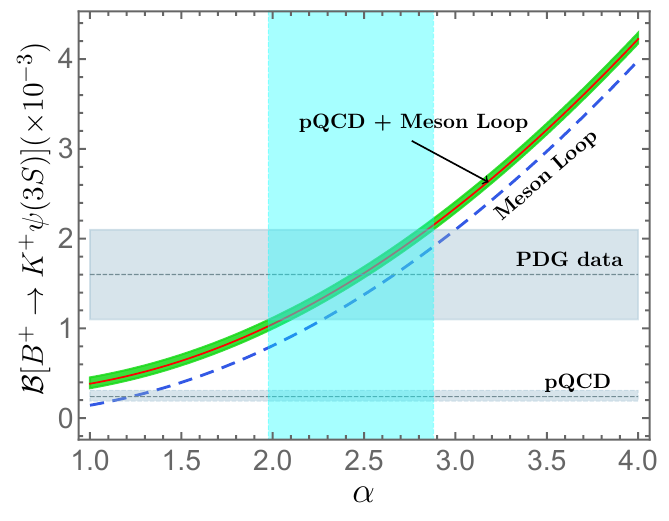}
	\caption{(Color online.) Branching fractions of $B^+ \rightarrow K^+\psi(4040)$ depending on the parameter $\alpha$.}
	\label{Fig:respsi}
\end{figure}

In Fig.~\ref{Fig:respsi}, we present our estimations for the branching fractions for $B^+ \to K^+ \psi(3S)$, where the horizontal dashed line with grey band is the short-range pQCD estimation and the blue dashed curve is the long-range contribution from the meson loops, which increase with the model parameter $\alpha$ . The red solid curve with grey band is the sum of short-range and long range contributions. In addition, the PDG average is also presented for comparison. From the figure, one can find that our estimations, considering both short range and long range contributions, can successfully reproduce the experimental data within the parameter range of $1.95 < \alpha < 2.88$ (cyan vertical band), which is strictly of order unity. In this parameter range, the meson loop contributions are essential for $B^+\to K^+ \psi(3S)$.

\begin{figure}[t]
	\centering
	\includegraphics[width=8.5cm]{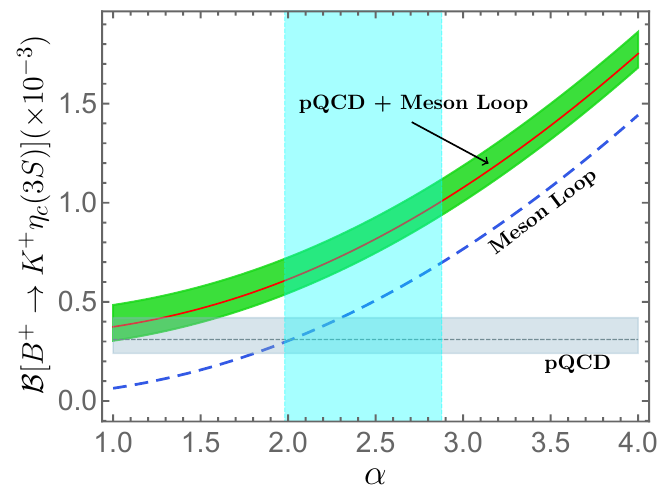}
	\caption{(Color online.) Branching fractions of $B^+ \rightarrow K^+\eta_c(4043) $ depending on the parameter $\alpha$.}
	\label{Fig:resetac}
\end{figure}

In the same way, we present the estimated branching fraction of $B^+ \to K^+ \eta_c(3S)$ in Fig.~\ref{Fig:resetac}. Within the parameter range determined by $B^+ \to K^+ \psi(3S)$, the branching fraction of $B^+ \to K^+ \eta_c(3S)$ is estimated  to be,
\begin{equation}
	\mathcal{B}[B^+ \to K^+ \eta_c(3S)]=(0.81_{-0.29}^{+0.31})\times 10^{-3}.
\end{equation}
where the central value is estimated with $\alpha=2.5$, and the uncertainties come from the pQCD estimations and the variation of $\alpha$. Unlike the $B^+ \to K^+ \psi(3S)$, the pQCD estimations and long-range meson loop contributions to the branching fraction of $B^+ \to K^+ \eta_c(3S)$ are similar in the parameter range determined by the branching fraction of $B^+ \to K^+ \psi(3S)$. Moreover, the established experimental branching fractions for the lower-lying $\eta_c$ states are reported to be~\cite{ParticleDataGroup:2026mpi},
\begin{eqnarray}
	\mathcal{B}(B^+ \rightarrow \eta_c K^+) &=& (1.09 \pm 0.08) \times 10^{-3}, \nonumber\\
	\mathcal{B}(B^+ \rightarrow \eta_c(2S) K^+) &=& (4.4 \pm 1.0) \times 10^{-4}.
\end{eqnarray}
As we can see, the central value of the predicted branching fraction of the $\eta_c(3S)$ is notably higher than the one of $\eta_c(2S)$, which is similar to the experimental observation in the vector sector, where the branching fraction of $B^+ \rightarrow K^+ \psi(3S) $ is larger than that of $B^+ \rightarrow \psi(2S) K^+$. Based on this estimation, we can conclude that the long-distance meson-loop mechanism plays a crucial role in accommodating the anomalous behavior of higher charmonium production in $B^+$ decay processes.

\begin{figure}
	\centering
	\includegraphics[width=8.5cm]{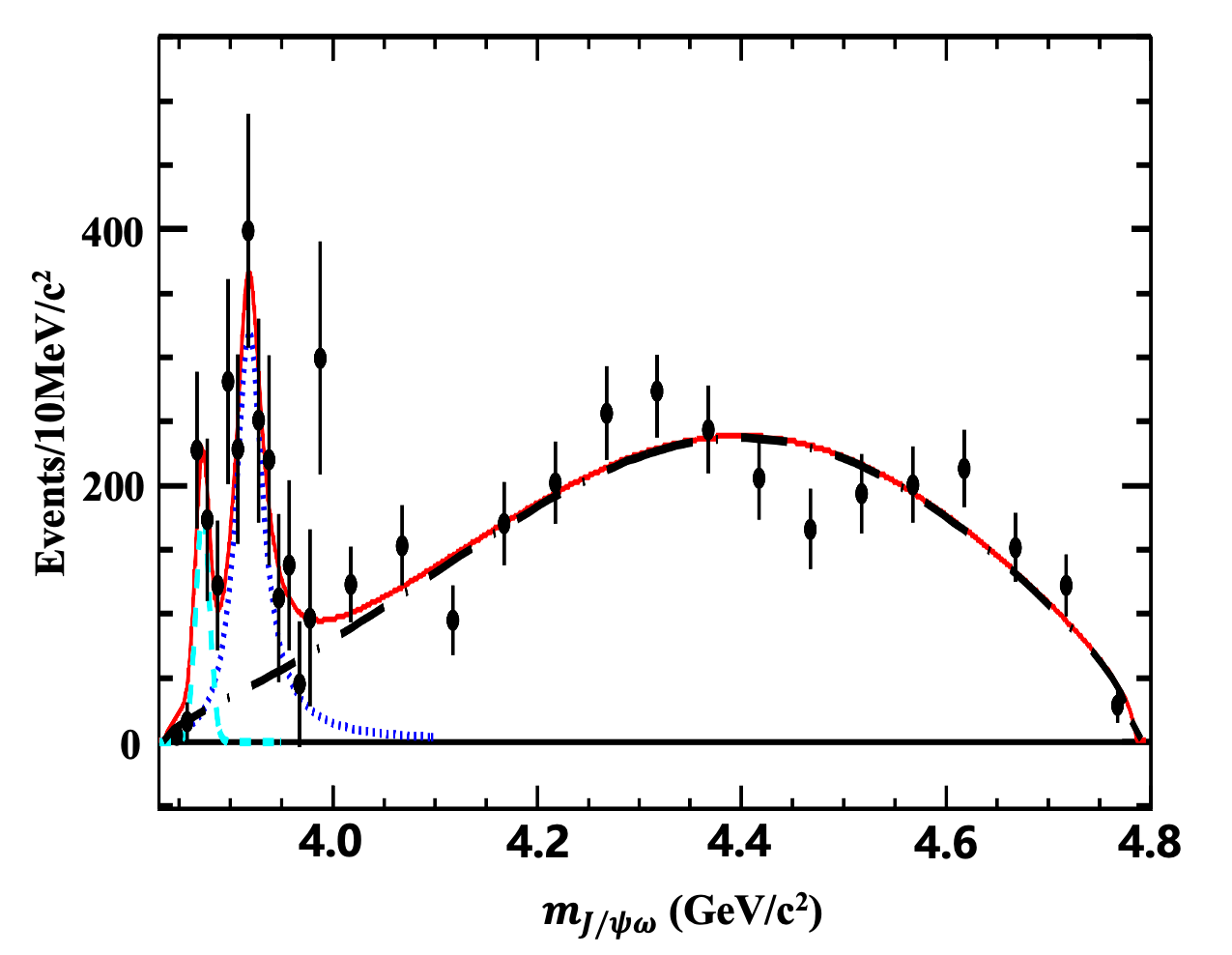}
	\caption{The corrected $m_{ J /\psi \omega}$ distribution for $B^+ \to J/\psi \omega K^+$ process in
    Ref.~\cite{BaBar:2010wfc}.}
	\label{Fig:BaBar}
\end{figure}

It is worth noting that the BaBar Collaboration reported their analysis of $B^+ \to J/\psi \omega K^+$ and $B^0 \to J/\psi \omega K^0$ in Ref.~\cite{BaBar:2010wfc}, and the $J/\psi \omega$ invariant mass distributions were measured. We present their mass distributions of $B^+ \to J/\psi \omega K^+$ process in Fig.~\ref{Fig:BaBar} for the convenience of readers. As shown in Fig.~\ref{Fig:BaBar}, there is an enhancement with one data point close to $4.0$ GeV in addition to the structures $X(3872)$ and $Y(3940)$, which may correspond to the structure of $\eta_c(3S)$, and $\eta_c(3S)$ further decays into $ J/\psi \omega$. With this evidence and the present estimation, we suggest that the experimentalist search for $\eta_c(3S)$ in the $J/\psi \omega$ invariant mass distributions of $B^+ \to J/\psi \omega K^+$ process, which should be accessible for the Belle II and LHCb Collaborations.

\section{SUMMARY}
\label{Sec:Sum}
Bottom meson decay is one of the most important charmonium and charmonium-like states production process. Both the short-distance perturbative and long-distance non-perturbative QCD effects are essential in understanding the higher charmonia production from bottom meson decay. In the present estimation, we systematically investigate the production of $\psi(3S)$ in $B^+ \to K^+ \psi(3S)$ by taking the short-distance perturbative and long-distance non-perturbative QCD effects into consideration, where the long-distance non-perturbative QCD effect is phenomenologically described by the meson loop contributions. Our estimations indicate that the branching fraction of $B^+ \to K^+\psi(3S)$ resulting from the short-range perturbative QCD estimations is several times smaller than the experimental measurements. After taking the long-distance non-perturbative effect into consideration, we find that the measured branching fraction of $B^+ \to K^+\psi(3S)$ can be well reproduced in a reasonable model parameter range. 

Considering the structural similarity of $\eta_c(3S)$ and $\psi(3S)$, we also investigate the production of $\eta_c(3S)$ in $B^+$ decay process by considering both the short-distance perturbative and long-distance non-perturbative QCD effects. Within the model parameter range determined by $B^+ \to K^+ \psi(3S)$, we predict the branching fraction of $B^+ \to K^+ \eta_c(3S)$ to be $(0.81_{-0.29}^{+0.31})\times 10^{-3}$, which is also greater than that of $B^+ \to K^+ \eta_c(2S)$. Based on the present estimations and the relevant experimental measurements, we suggest that the experimentalists search for $\eta_c(3S)$ in the $J/\psi \omega$ invariant mass distributions of $B^+ \to J/\psi \omega K^+$ process, which should be accessible for the Belle II and LHCb Collaborations.

\section*{ACKNOWLEDGMENTS}
This work is supported by the National Natural Science Foundation of China under the Grant Nos. 12335001 and 12675124, as well as supported, in part, by National Key Research and Development Program under Grant No.2024YFA1610503.

\begin{widetext}
\appendix
\section{pQCD factorization formulas }\label{Sec:AppendixC}
Explicit expressions for some functions appearing in the pQCD factorization formulas [Eqs.~(\ref{eq:f1})--(\ref{eq:m2})] are presented below.
We adopt the following model for the $B$ meson DA:
\[
\phi_B(x, b) = N_B x^2 (1 - x)^2 \exp \left[ -\frac{1}{2} \left( \frac{x m_B}{\omega_B} \right)^2 - \frac{1}{2} (\omega_B b)^2 \right],
\]
where $b$ is the impact parameter conjugate to the spectator transverse momentum $k_T$, $N_B$ is the normalization constant, and the shape parameter $\omega_B = 0.40\pm0.04$ GeV has been determined from studies of $B$ meson transition form factors~\cite{Keum:2000ph}.

We adopt the well-constrained kaon LCDAs from Ref.~\cite{Ali:2007ff},
\begin{eqnarray}
\phi_a(x) &=& \frac{3f_K}{\sqrt{6}} x(1 - x)[1 + 0.17 C_1^{3/2}(t) + 0.2 C_2^{3/2}(t)], \nonumber \\
\phi_p(x) &=& \frac{f_K}{2\sqrt{6}} [1 + 0.24 C_2^{1/2}(t)], \nonumber\\
\phi_t(x) &=& - \frac{f_K}{2\sqrt{6}} [C_1^{1/2}(t) + 0.35 C_3^{1/2}(t)],
\end{eqnarray}
where $f_K=0.16$ GeV and the Gegenbauer polynomials are defined as
\begin{eqnarray}
C_1^{1/2}(t) &=& t, \qquad C_1^{3/2}(t) = 3t, \qquad C_2^{1/2}(t) = \frac{1}{2}(3t^2 - 1), \nonumber \\
C_2^{3/2}(t) &=& \frac{3}{2}(5t^2 - 1), \qquad C_3^{1/2}(t) = \frac{1}{2}t(5t^2 - 3), 
\end{eqnarray}
with $t = 2x - 1$.

The LCDAs of the charmonium states for the $3S$ state  up to twist-3 have been derived in Ref.~\cite{Rui:2016opu}:
\begin{eqnarray}
\psi_{L}(x, b) &=& \frac{f_{\psi(3S)} }{2\sqrt{2N_c}} N^L x \bar{x} \mathcal{T}(x) e^{-x\bar{x}\frac{m_c}{\omega}\left[\omega^2 b^2 + \left(\frac{x-\bar{x}}{2x\bar{x}}\right)^2\right]}\quad , \nonumber \\
\psi_t(x, b) &=& \frac{f_{\psi(3S)}}{2\sqrt{2N_c}} N^t (x - \bar{x})^2 \mathcal{T}(x)   e^{-x\bar{x}\frac{m_c}{\omega}\left[\omega^2 b^2 + \left(\frac{x-\bar{x}}{2x\bar{x}}\right)^2\right]}\quad , \nonumber \\
\psi_v(x, b) &=& \frac{f_{\eta_c(3S)}}{2\sqrt{2N_c}} N^V [1 + (x - \bar{x})^2] \mathcal{T}(x)  e^{-x\bar{x}\frac{m_c}{\omega}\left[\omega^2 b^2 + \left(\frac{x-\bar{x}}{2x\bar{x}}\right)^2\right]}\quad , \nonumber \\
\psi_s(x, b) &=& \frac{f_{\eta_c(3S)}}{2\sqrt{2N_c}} N^s \mathcal{T}(x) e^{-x\bar{x}\frac{m_c}{\omega}\left[\omega^2 b^2 + \left(\frac{x-\bar{x}}{2x\bar{x}}\right)^2\right]} \quad ,  
\end{eqnarray}
where
\begin{eqnarray}
\mathcal{T}(x) &= 7 - 2 \frac{m_c (x - \bar{x})^2}{\omega x \bar{x}} - 24 m_c \omega b^2 x (1 - x) + \left( \frac{m_c (x - \bar{x})^2}{\omega x \bar{x}} - 4 b^2 m_c \omega x \bar{x} \right)^2.
\end{eqnarray}
with $m_c=1.275$ GeV and $\omega=0.1$ GeV~\cite{Rui:2016opu}.
The normalization constants $N^i$ are determined by  
\(\int_0^1 \psi_i(x, 0) dx = \frac{f_{3S}}{2\sqrt{2N_c}}\).
For the decay constants, we adopt the values $f_{\psi(3S)}=0.187$ GeV and  $f_{\eta_c(3S)}=0.18$ GeV from Ref.~\cite{Wang:2007fs}.
The hard scales are chosen as 
\begin{eqnarray}
t_{a,b} &= &\max(\sqrt{\beta_{a,b}}, 1/b_1, 1/b_3), \nonumber \\
t_{c,d} &= &\max(\sqrt{\alpha_e}, \sqrt{\beta_{c,d}}, 1/b_1, 1/b_2).
\end{eqnarray}
with
\begin{eqnarray}
\alpha_e &=& x_1 x_3 (1 - r^2) M^2, \qquad \beta_a = x_3 (1 - r^2) M^2, \nonumber\\
\beta_b &=& x_1 (1 - r^2) M^2, \nonumber\\
\beta_c &=& [(x_1 + x_2 - 1)(x_3 + r^2(1 - x_2 - x_3)) + r_c^2] M^2, \nonumber\\
\beta_d &=& [(x_1 - x_2)(x_3 + r^2(x_2 - x_3)) + r_c^2] M^2.  
\end{eqnarray}
The hard functions are defined by
\begin{eqnarray}
h(\alpha, \beta, b_1, b_2) &=&h_1(\alpha, b_1) \times h_2(\beta, b_1, b_2), \nonumber \\
h_1(\alpha, b_1) &= &\begin{cases} K_0(\sqrt{\alpha}b_1), & \alpha > 0, \\ K_0(i\sqrt{-\alpha}b_1), & \alpha < 0, \end{cases} \nonumber\\
h_2(\beta, b_1, b_2) &=&\begin{cases} \theta(b_1 - b_2) I_0(\sqrt{\beta}b_2) K_0(\sqrt{\beta}b_1) + (b_1 \leftrightarrow b_2), & \beta > 0, \\ \theta(b_1 - b_2) J_0(\sqrt{-\beta}b_2) K_0(i\sqrt{-\beta}b_1) + (b_1 \leftrightarrow b_2), & \beta < 0, \end{cases}
\end{eqnarray}
where $J_0$ is the Bessel function and $K_0$, $I_0$ are modified Bessel functions with $K_0(ix) = \frac{\pi}{2}(-N_0(x) + iJ_0(x))$.  

The Sudakov factors can be written as 
\begin{eqnarray}
S_{ab}(t) &=& s\left( \frac{M_B}{\sqrt{2}} x_1, b_1 \right) + s\left( \frac{M_B}{\sqrt{2}} x_3 (1 - r^2), b_3 \right)+ s\left( \frac{M_B}{\sqrt{2}} (1 - x_3)(1 - r^2), b_3 \right) \nonumber\\
&& + \frac{5}{3} \int_{1/b_1}^t \frac{\mathrm{d}\mu}{\mu} \gamma_q(\mu) + 2 \int_{1/b_3}^t \frac{\mathrm{d}\mu}{\mu} \gamma_q(\mu), \nonumber\\[10pt]
S_{cd}(t) &=& s\left( \frac{M_B}{\sqrt{2}} x_1, b_1 \right)  + s\left( \frac{M_B}{\sqrt{2}} x_2, b_2 \right) + s\left( \frac{M_B}{\sqrt{2}} (1 - x_2), b_2 \right)  + s\left( \frac{M_B}{\sqrt{2}} x_3 (1 - r^2), b_1 \right)   + s\left( \frac{M_B}{\sqrt{2}} (1 - x_3)(1 - r^2), b_1 \right) \nonumber \\
&& + \frac{11}{3} \int_{1/b_1}^t \frac{\mathrm{d}\mu}{\mu} \gamma_q(\mu) + 2 \int_{1/b_2}^t \frac{\mathrm{d}\mu}{\mu} \gamma_q(\mu),
\end{eqnarray}
with $\gamma_q = -\alpha_s/\pi$ being the quark anomalous dimension. The function $s(Q, b)$ is expressed as
\begin{eqnarray}
s(Q, b) = \frac{A^{(1)}}{2\beta_1} \hat{q} \ln\left(\frac{\hat{q}}{\hat{b}}\right) - \frac{A^{(1)}}{2\beta_1} (\hat{q} - \hat{b}) + \frac{A^{(2)}}{4\beta_1^2} \hat{q} \ln\left(\frac{\hat{q}}{\hat{b}} - 1\right) - \left[ \frac{A^{(2)}}{4\beta_1^2} - \frac{A^{(1)}}{4\beta_1} \ln\left(\frac{e^{2\gamma_E - 1}}{2}\right) \right] \ln\left(\frac{\hat{q}}{\hat{b}}\right),
\end{eqnarray}
where
\begin{eqnarray}
\hat{q} = \ln \frac{Q}{\sqrt{2}\Lambda_{\text{QCD}}} , \qquad \hat{b} = \ln \frac{1}{b\Lambda_{\text{QCD}}} , \qquad \beta_1 = \frac{33 - 2n_f}{12} , \qquad \beta_2 = \frac{153 - 19n_f}{24} ,
\end{eqnarray}
\begin{eqnarray}
A^{(1)} = \frac{4}{3} , \qquad A^{(2)} = \frac{67}{9} - \frac{\pi^2}{3} - \frac{10}{27}n_f + \frac{8}{3}\beta_1 \ln\left(\frac{1}{2}e^{\gamma_E}\right),
\end{eqnarray}
with $n_f$ being the number of the quark flavor, and $\gamma_E$ the Euler constant.
The threshold resummation factor $S_t (x)$ is adopted from~\cite{Kurimoto:2001zj}
\begin{eqnarray}
S_t(x) = \frac{2^{1+2c} \Gamma(3/2 + c)}{\sqrt{\pi} \Gamma(1 + c)} [x(1 - x)]^c
\end{eqnarray}
with $c=0.3$.

\section{Decay amplitudes of $B^+ \to K^+ \psi(3S)$}\label{Sec:AppendixA}
The amplitudes corresponding to the loops contributing to process $B^+ \to K^+ \psi(3S) $ are
\begin{eqnarray}
    \begin{split}
    \mathcal{M}_{(D_s^{+} \bar{D}^{0}) D^{*0}}&= i^{3} \int\frac{d^{4}q}{(2\pi)^{4}}\Big[\mathcal{A}^{B\rightarrow D_{s} \bar{D}^{0}}(p_1,p_2)\Big]\Big[-g_{D^{*}DP}(i p_{3\beta})\Big]\Big[i g_{\psi D^{*}D}\varepsilon_{\mu_{1}\delta\phi\beta_1}p_4^{\mu_1}(p_2-q)^{\beta_1}\epsilon^{\delta}(p_4)\Big]\\
    &\times\Big[\frac{1}{p_1^{2}-m_1^{2}}\Big]\Big[\frac{1}{p_2^{2}-m_2^{2}}\Big]\Big[\frac{-g^{\beta\phi}+q^{\beta} q^{\phi}/m_q^{2}}{q^{2}-m_q^{2}}\Big]\mathcal{F}^{2}(q^{2},m_q^{2})\\
    \mathcal{M}_{(D_s^{*+} \bar{D}^{0}) D^{0}}&= i^{3} \int\frac{d^{4}q}{(2\pi)^{4}}\Big[\mathcal{A}_\mu^{B\rightarrow D_{s}^{*} \bar{D}^{0}}(p_1,p_2)\Big]\Big[g_{D^{*}DP}(i p_{3\alpha})\Big]\Big[i g_{\psi DD}(q-p_2)^{\delta}\epsilon_\delta(p_4)\Big]\\
    &\times\Big[\frac{-g^{\mu\alpha}+p_1^{\mu} p_1^{\alpha}/m_1^{2}}{p_1^{2}-m_1^{2}}\Big]\Big[\frac{1}{p_2^{2}-m_2^{2}}\Big]\Big[\frac{1}{q^{2}-m_q^{2}}\Big]\mathcal{F}^{2}(q^{2},m_q^{2})\\
    \mathcal{M}_{(D_s^{*+} \bar{D}^{0}) D^{*0}}&= i^{3} \int\frac{d^{4}q}{(2\pi)^{4}}\Big[\mathcal{A}_\mu^{B\rightarrow D_{s}^{*} \bar{D}^{0}}(p_1,p_2)\Big]\Big[\frac{i}{2} g_{D^{*}D^{*}P}\varepsilon_{\beta\nu_1\alpha_1\alpha}p_3^{\nu_1}(p_1+q)^{\alpha_1}\Big]\Big[i g_{\psi D^{*}D}\varepsilon_{\mu_1\delta\phi\beta_1}p_4^{\mu_1}(p_2-q)^{\beta_1}\epsilon^{\delta}(p_4)\Big]\\
    &\times\Big[\frac{-g^{\mu\alpha}+p_1^{\mu} p_1^{\alpha}/m_1^{2}}{p_1^{2}-m_1^{2}}\Big]\Big[\frac{1}{p_2^{2}-m_2^{2}}\Big]\Big[\frac{-g^{\beta\phi}+q^{\beta} q^{\phi}/m_q^{2}}{q^{2}-m_q^{2}}\Big]\mathcal{F}^{2}(q^{2},m_q^{2})\\
    \mathcal{M}_{(D_s^{+} \bar{D}^{*0}) D^{*0} }&= i^{3} \int\frac{d^{4}q}{(2\pi)^{4}}\Big[\mathcal{A}_\nu^{B\rightarrow D_{s}\bar{D}^{*0}}(p_1,p_2)\Big]\Big[- g_{D^{*}DP}(i p_{3\beta})\Big]\Big[i g_{\psi D^{*}D^{*}}g^{\mu_1\delta}\epsilon_\delta(p_4)\big(g_{\nu_1\phi}g_{\mu_1\rho}(p_2-q)^{\nu_1}+g_{\mu_1\phi}g_{\nu_1\rho}(p_2-q)^{\nu_1}\\
    &-g_{\nu_1\phi}g^{\nu_1\rho}(p_2-q)_{\mu_1}\big)\Big]\Big[\frac{1}{p_1^{2}-m_1^{2}}\Big]\Big[\frac{-g^{\nu\rho}+p_2^{\nu} p_2^{\rho}/m_2^{2}}{p_2^{2}-m_2^{2}}\Big]\Big[\frac{-g^{\beta\phi}+q^{\beta} q^{\phi}/m_q^{2}}{q^{2}-m_q^{2}}\Big]\mathcal{F}^{2}(q^{2},m_q^{2})\\
    \mathcal{M}_{(D_s^{*+} \bar{D}^{*0}) D^{0}}&= i^{3} \int\frac{d^{4}q}{(2\pi)^{4}}\Big[\mathcal{A}_{\mu\nu}^{B\rightarrow D_{s}^{*} \bar{D}^{*0}}(p_1,p_2)\Big]\Big[ g_{D^{*}D^{*}P}(i p_{3\alpha})\Big]\Big[-i g_{\psi D^{*}D}\varepsilon_{\mu_1\delta\rho\beta_1}p_4^{\mu_1}(p_2-q)^{\beta_1}\epsilon^{\delta}(p_4)\Big]\\
    &\times\Big[\frac{-g^{\mu\alpha}+p_1^{\mu} p_1^{\alpha}/m_1^{2}}{p_1^{2}-m_1^{2}}\Big]\Big[\frac{-g^{\nu\rho}+p_2^{\nu} p_2^{\rho}/m_2^{2}}{p_2^{2}-m_2^{2}}\Big]\Big[\frac{1}{q^{2}-m_q^{2}}\Big]\mathcal{F}^{2}(q^{2},m_q^{2})\\
    \mathcal{M}_{(D_s^{*+} \bar{D}^{*0}) D^{*0}}&= i^3 \int\frac{d^{4}q}{(2\pi)^{4}}\Big[\mathcal{A}_{\mu\nu}^{B\rightarrow D_{s}^{*}\bar{D}^{*0}}(p_1,p_2)\Big]\Big[\frac{i}{2} g_{D^{*}D^{*}P}\varepsilon_{\beta\nu_1\alpha_1\alpha}p_3^{\nu_1}(p_1+q)^{\alpha_1}\Big]\Big[i g_{\psi D^{*}D^{*}}g^{\mu_1\delta}\epsilon_\delta(p_4)\big(g_{\nu_1\phi}g_{\mu_1\rho}(p_2-q)^{\nu_1}\\
    &+g_{\mu_1\phi}g_{\nu_1\rho}(p_2-q)^{\nu_1}-g_{\nu_1\phi}g^{\nu_1\rho}(p_2-q)_{\mu_1}\big)\Big]\Big[\frac{-g^{\mu\alpha}+p_1^{\mu} p_1^{\alpha}/m_1^{2}}{p_1^{2}-m_1^{2}}\Big]\Big[\frac{-g^{\nu\rho}+p_2^{\nu} p_2^{\rho}/m_2^{2}}{p_2^{2}-m_2^{2}}\Big]\Big[\frac{-g^{\beta\phi}+q^{\beta} q^{\phi}/m_q^{2}}{q^{2}-m_q^{2}}\Big]\mathcal{F}^{2}(q^{2},m_q^{2})\\
    \mathcal{M}_{(D_{s1}^{+} \bar{D}^{*0}) D^{*0}}&= i^{3} \int\frac{d^{4}q}{(2\pi)^{4}}\Big[\mathcal{A}_{\mu\nu}^{B\rightarrow D_{s1}^{+} \bar{D}^{*0}}(p_1,p_2)\Big]\Big[i g_{D^{*} D_1 P}(i (q+p_1)_\alpha(i p_{3\beta})-i(q+p_1)^{\alpha_2}g_{\alpha\beta}i p_{3\alpha_2}+i(q+p_1)_\beta i p_{3\alpha})\Big]\\
    &\times \Big[i g_{\psi D^{*}D^{*}} g^{\mu_3 \delta}\epsilon_\delta(p_4) (g_{\nu_3 \phi}g_{\rho \mu_3}(p_2-q)^{\nu_3}+g_{\mu_3 \phi}g_{\nu_3 \rho}(p_2-q)^{\nu_3}-g_{\nu_3 \phi}g_{\nu_3\rho}(p_2-q)^{\mu_3}\Big]\Big[\frac{-g^{\mu\alpha}+p_1^{\mu} p_1^{\alpha}/m_1^{2}}{p_1^{2}-m_1^{2}}\Big]\\
    &\times\Big[\frac{-g^{\nu\rho}+p_2^{\nu} p_2^{\rho}/m_2^{2}}{p_2^{2}-m_2^{2}}\Big] \Big[\frac{-g^{\beta\phi}+q^{\beta} q^{\phi}/m_q^{2}}{q^{2}-m_q^{2}}\Big]\mathcal{F}^{2}(q^{2},m_q^{2})\\
    \mathcal{M}_{(D_{s1}^{+} \bar{D}^{0}) D^{*0}}&= i^{3} \int\frac{d^{4}q}{(2\pi)^{4}}\Big[\mathcal{A}_{\mu\nu}^{B\rightarrow D_{s1}^{+} \bar{D}^{*0}}(p_1,p_2)\Big]\Big[i g_{D^{*} D_1 P}(i (q+p_1)_\alpha(i p_{3\beta})-i(q+p_1)^{\alpha_2}g_{\alpha\beta}i p_{3\alpha_2}+i(q+p_1)_\beta i p_{3\alpha})\Big]\\
    &\times \Big[-i g_{\psi D^{*}D}\varepsilon_{\mu_1\delta\rho\beta_1}p_{4}^{\mu_1}(p_2-q)^{\beta_1}\epsilon^{\delta}(p_4)\Big]\Big[\frac{-g^{\mu\alpha}+p_1^{\mu} p_1^{\alpha}/m_1^{2}}{p_1^{2}-m_1^{2}}\Big]\Big[\frac{-g^{\nu\rho}+p_2^{\nu} p_2^{\rho}/m_2^{2}}{p_2^{2}-m_2^{2}}\Big] \Big[\frac{-g^{\beta\phi}+q^{\beta} q^{\phi}/m_q^{2}}{q^{2}-m_q^{2}}\Big]\mathcal{F}^{2}(q^{2},m_q^{2})\\
    \mathcal{M}_{(\bar{D}^{*0} D_{s}^{+} ) D_s^{*-}}&= i^3 \int\frac{d^{4}q}{(2\pi)^{4}}\Big[\mathcal{A}_{\mu}^{B\rightarrow \bar{D}^{0} D_{s}^{+}}(p_1,p_2)\Big]\Big[-g_{D^{*} D P} i p_{3\alpha} \Big]\Big[-i g_{\psi D_s D_s} \epsilon_\delta(p_4)(q-p_2)^{\delta }\Big]\Big[\frac{-g^{\mu\alpha}+p_1^{\mu} p_1^{\alpha}/m_1^{2}}{p_1^{2}-m_1^{2}}\Big]\mathcal{F}^{2}(q^{2},m_q^{2})\\
    \end{split}
\end{eqnarray}

\section{Decay amplitudes of $B^+ \to K^+ \eta_c(3S)$}\label{Sec:AppendixB}
\begin{figure*}[t]
	\begin{tabular}{ccc}
		\centering
		\includegraphics[width=4.8cm]{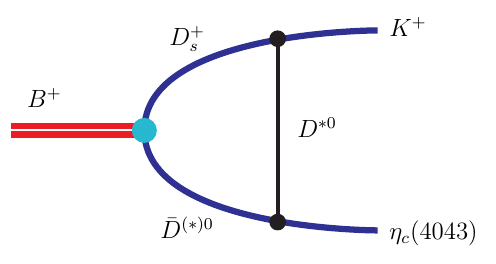}&
		\includegraphics[width=4.8cm]{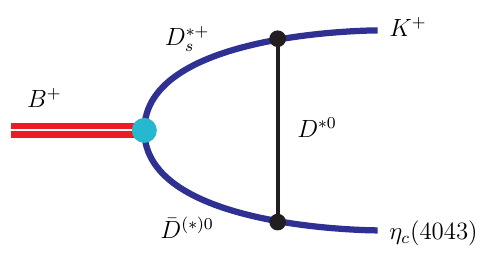}&
		\includegraphics[width=4.8cm]{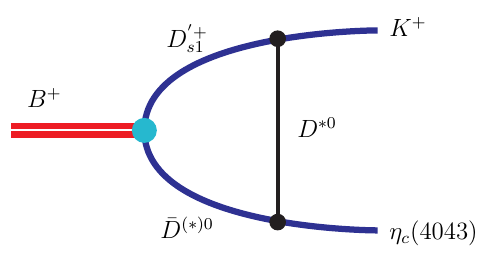}\\
		$(e)$ & $(f)$ & $(g)$ \\ \\
	\end{tabular}
	\caption{Diagrams contributing to $B^+ \to K^+ \eta_c(4043)$ at the hadron level.}\label{fig:FeynDiagram2}
\end{figure*}

The amplitudes corresponding to the loops contributing to process $B^+ \to K^+ \eta_c(3S) $ are
\begin{eqnarray}
    \begin{split}
    \mathcal{M}_{(D_s^{+} \bar{D}^{0}) D^{*0}}&= i^{3} \int\frac{d^{4}q}{(2\pi)^{4}}\Big[\mathcal{A}^{B\rightarrow D_{s} \bar{D}^{0}}(p_1,p_2)\Big]\Big[-g_{D^{*}DP}(i p_{3\beta})\Big]\Big[-i g_{\eta_c D^* D}(p_2-q)_\phi \Big]\\
    &\times\Big[\frac{1}{p_1^{2}-m_1^{2}}\Big]\Big[\frac{1}{p_2^{2}-m_2^{2}}\Big]\Big[\frac{-g^{\beta\phi}+q^{\beta} q^{\phi}/m_q^{2}}{q^{2}-m_q^{2}}\Big]\mathcal{F}^{2}(q^{2},m_q^{2})\\
    \mathcal{M}_{(D_s^{*+} \bar{D}^{0}) D^{*0}}&= i^{3} \int\frac{d^{4}q}{(2\pi)^{4}}\Big[\mathcal{A}_\mu^{B\rightarrow D_{s}^{*} \bar{D}^{0}}(p_1,p_2)\Big]\Big[\frac{i}{2} g_{D^{*}D^{*}P}\varepsilon_{\beta\nu_1\alpha_1\alpha}p_3^{\nu_1}(p_1+q)^{\alpha_1}\Big]\Big[-i g_{\eta_c D^* D}(p_2-q)_\phi]\\
    &\times\Big[\frac{-g^{\mu\alpha}+p_1^{\mu} p_1^{\alpha}/m_1^2}{p_1^2-m_1^2}\Big]\Big[\frac{1}{p_2^2-m_2^2}\Big]\Big[\frac{-g^{\beta\phi}+q^{\beta} q^{\phi}/m_q^2}{q^2-m_q^2}\Big]\mathcal{F}^2(q^2,m_q^2)\\
    \mathcal{M}_{(D_s^+ \bar{D}^{*0}) D^{*0} }&= i^3 \int\frac{d^4q}{(2\pi)^4}\Big[\mathcal{A}_\nu^{B\rightarrow D_{s}\bar{D}^{*0}}(p_1,p_2)\Big]\Big[- g_{D^*DP}(i p_{3\beta})\Big]\Big[-i g_{\eta_c D^{*} D^{*}}\epsilon_{\mu_3 \phi \alpha_3 \rho}(i p_4^{\mu_3})(-i)(p_2-q)^{\alpha_3} \Big]\\
    &\times \Big[\frac{1}{p_1^2-m_1^2}\Big]\Big[\frac{-g^{\nu\rho}+p_2^{\nu} p_2^{\rho}/m_2^2}{p_2^2-m_2^2}\Big]\Big[\frac{-g^{\beta\phi}+q^{\beta} q^{\phi}/m_q^2}{q^2-m_q^2}\Big]\mathcal{F}^2(q^2,m_q^2)\\
    \mathcal{M}_{(D_s^{*+} \bar{D}^{*0}) D^0}&= i^3 \int\frac{d^4q}{(2\pi)^4}\Big[\mathcal{A}_{\mu\nu}^{B\rightarrow D_{s}^* \bar{D}^{*0}}(p_1,p_2)\Big]\Big[ g_{D^{*}D^{*}P}(i p_{3\alpha})\Big]\Big[-i g_{\eta_c D^{*} D}(p_2-q)_\rho \Big]\\
    &\times\Big[\frac{-g^{\mu\alpha}+p_1^{\mu} p_1^{\alpha}/m_1^2}{p_1^2-m_1^2}\Big]\Big[\frac{-g^{\nu\rho}+p_2^{\nu} p_2^{\rho}/m_2^2}{p_2^2-m_2^2}\Big]\Big[\frac{1}{q^2-m_q^2}\Big]\mathcal{F}^2(q^2,m_q^2)\\
    \mathcal{M}_{(D_s^{*+} \bar{D}^{*0}) D^{*0}}&= i^3 \int\frac{d^4q}{(2\pi)^4}\Big[\mathcal{A}_{\mu\nu}^{B\rightarrow D_{s}^{*}\bar{D}^{*0}}(p_1,p_2)\Big]\Big[\frac{i}{2} g_{D^{*}D^{*}P}\varepsilon_{\beta\nu_1\alpha_1\alpha}p_3^{\nu_1}(p_1+q)^{\alpha_1}\Big]\Big[-i g_{\eta_c D^{*} D^{*}}\epsilon_{\mu_3 \phi \alpha_3 \rho}(i p_4^{\mu_3})(-i)(p_2-q)^{\alpha_3} \Big]\\
    &\times \Big[\frac{-g^{\mu\alpha}+p_1^{\mu} p_1^{\alpha}/m_1^2}{p_1^2-m_1^2}\Big]\Big[\frac{-g^{\nu\rho}+p_2^{\nu} p_2^{\rho}/m_2^2}{p_2^2-m_2^2}\Big]\Big[\frac{-g^{\beta\phi}+q^{\beta} q^{\phi}/m_q^2}{q^2-m_q^2}\Big]\mathcal{F}^2(q^2,m_q^2)\\
    \mathcal{M}_{(D_{s1}^{+} \bar{D}^{*0}) D^{*0}}&= i^3 \int\frac{d^4q}{(2\pi)^4}\Big[\mathcal{A}_{\mu\nu}^{B\rightarrow D_{s1}^{+} \bar{D}^{*0}}(p_1,p_2)\Big]\Big[i g_{D^* D_1 P}(i (q+p_1)_\alpha(i p_{3\beta})-i(q+p_1)^{\alpha_2}g_{\alpha\beta}i p_{3\alpha_2}+i(q+p_1)_\beta i p_{3\alpha})\Big]\\
    &\times \Big[-i g_{\eta_c D^* D^*}\epsilon_{\mu_3 \phi \alpha_3 \rho}(i p_4^{\mu_3})(-i)(p_2-q)^{\alpha_3}\Big]\Big[\frac{-g^{\mu\alpha}+p_1^{\mu} p_1^{\alpha}/m_1^2}{p_1^2-m_1^2}\Big]\Big[\frac{-g^{\nu\rho}+p_2^{\nu} p_2^{\rho}/m_2^2}{p_2^2-m_2^2}\Big] \Big[\frac{-g^{\beta\phi}+q^{\beta} q^{\phi}/m_q^2}{q^2-m_q^2}\Big]\mathcal{F}^2(q^2,m_q^2)\\
    \mathcal{M}_{(D_{s1}^{+} \bar{D}^{0}) D^{*0}}&= i^3 \int\frac{d^4q}{(2\pi)^4}\Big[\mathcal{A}_{\mu\nu}^{B\rightarrow D_{s1}^{+} \bar{D}^{*0}}(p_1,p_2)\Big]\Big[i g_{D^{*} D_1 P}(i (q+p_1)_\alpha(i p_{3\beta})-i(q+p_1)^{\alpha_2}g_{\alpha\beta}i p_{3\alpha_2}+i(q+p_1)_\beta i p_{3\alpha})\Big]\\
    &\times \Big[-i g_{\eta_c D^{*} D}(p_2-q)_\phi \Big]\Big[\frac{-g^{\mu\alpha}+p_1^{\mu} p_1^{\alpha}/m_1^2}{p_1^2-m_1^2}\Big]\Big[\frac{1}{p_2^2-m_2^2}\Big] \Big[\frac{-g^{\beta\phi}+q^{\beta} q^{\phi}/m_q^2}{q^2-m_q^2}\Big]\mathcal{F}^2(q^2,m_q^2)\\
    \end{split}
\end{eqnarray}
\end{widetext}

\bibliographystyle{apsrev4-1}
\bibliography{B2Kpsi4040.bib}
\end{document}